\documentclass[%
 reprint,
 amsmath,amssymb,
 aps,
 nofootinbib
]{revtex4-2}

\usepackage{graphicx}
\usepackage{dcolumn}
\usepackage{bm}
\usepackage{physics} 
\usepackage{enumerate} 
\usepackage[hidelinks]{hyperref} 
\usepackage{tikz}
\usepackage{xcolor} 
\renewcommand{\div}{\nabla \cdot }
\renewcommand{\grad}{\nabla}

\begin{document}
\preprint{APS/123-QED}
\title{Hydrodynamic theory of wetting by active particles}

\author{Noah Grodzinski$^{1}$}\thanks{Corresponding author: njbg2@cam.ac.uk (he/him)}
\author{Michael E. Cates$^{1}$}
\author{Robert L. Jack$^{1,2}$}

\affiliation{$^1$Department of Applied Mathematics and Theoretical Physics, University of Cambridge, Wilberforce Road, Cambridge, United Kingdom}
\affiliation{$^2$ Yusuf Hamied Department of Chemistry, University of Cambridge, Lensfield Road, Cambridge, United Kingdom}
\date{\today}
\begin{abstract}

The accumulation of self-propelled particles on repulsive barriers is a widely observed feature in active matter. Despite being implicated in a broad range of biological processes, from biofilm formation to cytoskeletal movement, wetting of surfaces by active particles remains poorly understood. In this work, we study this active wetting by considering a model comprising an active lattice gas, interacting with a permeable barrier under periodic boundary conditions, for which an exact hydrodynamic description is possible. We consider a hydrodynamic scaling limit that eliminates dynamical noise while retaining microscopic interpretability, enabling a precise characterisation of steady-states and their transitions. We demonstrate that the accumulation of active particles has remarkable similarities to equilibrium wetting, and that active wetting transitions display all the salient characteristics of the equilibrium critical wetting transition -- despite fundamental differences in underlying microscopic dynamics. However, our framework also enables the investigation of subtle but important nonequilibrium effects in active wetting, including a spontaneous ratchet effect which leads to a global steady-state current, departure of the bulk densities from their binodal values, and a novel dynamical transition pathway. Our results provide an intrinsically nonequilibrium framework in which to study active wetting, precisely demonstrating the connection to passive wetting while clarifying the nonequilibrium consequences of activity. 

\end{abstract}

\maketitle
\section{Introduction}

Active matter comprises a broad class of self-propelled particles, which consume energy at a local scale to produce directed motion. Such particles are extremely common in microbiological and ecological contexts, spanning length scales from microns to meters and involving particle numbers from dozens to millions \cite{Marchetti2013, Ramaswamy2010, Elgeti2015}. Examples include swimming bacteria \cite{Tailleur2008, Cates2012, Alert2020}, flocking birds \cite{Vicsek1995, Toner1998}, and cytoskeletal networks \cite{Ramaswamy2010, Doostmohammadi2018} -- as well as synthetic cases such as Janus particles \cite{Walther2013} and robotic swarms \cite{Saintyves2024}.

The continuous consumption of energy drives active particles out of equilibrium, leading to a diverse range of emergent collective behaviours. For example, active particles undergo motility-induced phase separation (MIPS) into dense (liquid) and dilute (vapour) phases due to the interplay between volume exclusion and self-propulsion, even in the absence of explicitly attractive interactions~\cite{Cates2015, Cates2025}. It is expected that collective behaviour in active systems should show some degree of independence of the microscopic details of particle interactions \cite{Cates2013, Ramaswamy2010}, so simple particle-level \cite{Fily2012, Redner2013, Digregorio2018} or field-theoretical \cite{Wittkowski2014, Tjhung2018, Cates2025} models of active matter are natural settings for studying their emergent behaviours.

A distinguishing feature of active systems, compared to their passive counterparts, is the disproportionate influence of boundaries. Non-local effects arising from steady-state currents mean that boundaries play a much more central role in determining the collective behaviour of active matter than in the passive case. Indeed, recent studies have shown that boundaries can regulate transport, transfer energy between length scales, and create or destroy large-scale structure \cite{Granek2024, Bechinger2016, Stenhammar2016, BenDor2022, Galajda2007, Reichhardt2017, Mangeat2024}.

One prominent boundary-driven effect is the accumulation of active particles on repulsive barriers. Recent works have shown that the self-propulsion of active particles results in a layer of {excess} density on any repulsive wall or barrier \cite{Turci2021, Turci2024, Neta2021, Elgeti2013, Zhao2026, Granek2024, Caprini2024, Sepulveda2017, Sepulveda2018, Das2020, Das2018, Mangeat2024, Das2025, PerezBastias2025, Wysocki2020, RojasVega2023}; this behaviour is reminiscent of the wetting of surfaces in equilibrium \cite{DeGennes1985, Bonn2009, Evans2019}, and so the phenomenon has become known as active wetting {\cite{Thiele2026}}. Experiments have implicated active wetting in a wide variety of biological processes \cite{Tayar2023, Douezan2011}, including biofilm formation~\cite{Hartmann2018, Lauga2006, Yaman2019, Chai2024, Harshey2003}, epithelial tissue organisation~\cite{PerezGonzalez2018, Brugues2014}, and cytoskeletal movement~\cite{Joanny2013}. 

{In equilibrium systems, adsorption (a microscopically thin film of fluid near a surface) is carefully distinguished from genuine wetting transitions \cite{Binder1988, DeGennes1985, Sullivan1986, Dietrich1988, Schick1990}. In particular, equilibrium critical wetting is a continuous surface phase transition on varying a surface attraction parameter at two-phase coexistence, at which the adsorbed film thickness diverges \cite{Lipowsky1982}. A complete wetting transition involves the divergence of the wetting film approaching bulk coexistence on the wet side of the wetting transition. Simulations have suggested that in systems undergoing MIPS, accumulation/adsorption of active particles \cite{Sepulveda2017, Sepulveda2018, Das2018, Elgeti2013} can develop into macroscopic wetting \cite{Turci2024, Zhao2026}, with transitions analogous to equilibrium complete wetting \cite{Neta2021} and critical wetting \cite{Turci2021, Das2020, PerezBastias2025, Das2025}. }

Despite its significant role in active systems, and biological relevance, the physics of active wetting remains poorly understood -- complex interfacial phenomena and non-local effects mean that even basic results of equilibrium wetting theory are not necessarily applicable to the active case~\cite{Zakine2020, Zhao2026, Bialke2015, Turci2024, Adkins2022, Granek2024}. In particular, the phenomenology of an active interface cannot be rationalised in terms of a single effective surface tension~\cite{Bialke2015, Cates2025}. {These confounding effects also make it difficult to distinguish between the distinct phenomena of adsorption and critical/complete wetting -- these distinctions have not always been maintained in the active wetting literature \cite{Thiele2026, Granek2024, Bechinger2016}.} An intrinsically nonequilibrium framework is therefore required to understand these nonequilibrium surface phenomena. 

In this work, we study active wetting via a new approach, by analysing a lattice model of self-propelled particles, whose large-scale (hydrodynamic) behaviour can be derived exactly~\cite{KourbaneHoussene2018, Mason2023a, Erignoux2024}, and is governed by deterministic equations. A consequence of the hydrodynamic limit as taken here is that the large-scale behaviour of our model is not affected by fluctuation corrections (in the sense of the renormalisation group).  This leads to critical behaviour of mean-field type.  Nevertheless, our results provide generic insights for active wetting.

The hydrodynamic framework -- which has also provided important recent insights in other types of active system~\cite{Martin2025, Mason2025} -- allows us to use precise analytic techniques familiar from field theories, while retaining a direct connection to microscopic behaviour\footnote{One aspect that aids this connection is that particle density and orientation fields both appear in the hydrodynamic equations; this differs from most active field theories for isotropic fluids, which take the hydrodynamic limit in a different way that eliminates orientational dynamics but retains diffusive noise.}. This means we can identify subtle yet physically significant effects, which may be obscured in particle-level studies \cite{Turci2021} by the confounding role of noise.

In equilibrium systems, wetting is often investigated in a \textit{slit geometry}, wherein a fluid is confined between planar walls \cite{Evans2019, Leeuwen1989, Bucior2009, Swain2000, Valencia2001, Maciolek2003, Grzelak2008, Nijmeijer1990, Sikkenk1987}, simplifying the investigation of wetting transitions (see Section \ref{sec:background} and {Appendix \ref{app:wetting}}). 
We seek an analogous geometry to study active wetting transitions. However, it is known that active particles generically adsorb to hard walls \cite{Das2018, Elgeti2013, Sepulveda2017, Sepulveda2018}, leading to full wetting  only~\cite{Zhao2026, Turci2021, Das2020, Neta2021, Turci2024}. We therefore use a smooth, permeable barrier (as in \cite{Turci2021, Das2020}) to access partial wetting states. We take periodic boundaries to suppress any other boundary-driven phenomena, and use fixed number of particles (see Section \ref{subsec:active_dynamics}) such that the model is always at liquid-vapour coexistence {(the transition we observe is therefore analogous to \textit{critical}, rather than \textit{complete}, wetting)}. Our deterministic equations result in a system which remains translationally symmetric in the direction(s) parallel to the wall, and so we study one-dimensional ($1d$) profiles -- although our hydrodynamic model admits a general and exact description even in higher dimensions.

Using this framework, we demonstrate remarkable similarities between active and equilibrium wetting\footnote{The analogy with equilibrium wetting transitions requires that the barrier-free system is in a state of two-phase coexistence~\cite{Turci2021, Turci2024, Zhao2026} which is the case in this work.  Adsorption of active particles onto a boundary from a homogeneous fluid phase~\cite{Das2018, Elgeti2013, Sepulveda2017, Sepulveda2018} is sometimes also called active wetting in the literature, but this is not our focus here.}. Specifically, we show that both full- and partial-wetting states arise as a consequence of activity, separated by a critical wetting transition. These observations are surprising because active wetting takes place without any explicitly attractive particle-particle or particle-wall interactions, so connections to equilibrium liquids are not at all guaranteed. Our hydrodynamic framework therefore enables us to generalise the already powerful analogy between equilibrium phase separation and active MIPS to include surface phenomena. 

Our model also enables the identification of subtle but important nonequilibrium phenomena in active wetting, which arise as a consequence of activity. In particular, we demonstrate that a ratchet effect emerges on a partially wet barrier, leading to steady-state currents (as may be generically expected in nonequilibrium steady-states~\cite{Turci2024, Zhao2026, BenDor2022, Granek2024}). The ratchet current leads to a number of intrinsically nonequilibrium effects in active wetting, including departure of the bulk phases from their binodal densities and the emergence of a novel dynamical pathway for the wetting transition -- these are explored in the companion Letter \cite{Grodzinski2026a}. However, the scaling laws associated with critical wetting remain unchanged. 
Particle-based simulations also indicated that active wetting is critical in $2d$, but large fluctuation effects made this hard to characterise~\cite{Turci2021}.  Persistent currents were also found previously for active wetting in a different (droplet) geometry, where they affect the contact angle via a correction to Young's equation~\cite{Zhao2026}. 
Our results complement these previous works, in particular the combination of the hydrodynamic approach and slit geometry lead to efficient characterisation of the wetting transition, without confounding finite-size or fluctuation effects. Overall, we find active wetting states to display an equilibrium-like critical-wetting surface phase transition, but one which is also modified by the emergence of spontaneous nonequilibrium current flow.  

The remainder of this paper is structured as follows: in Section \ref{sec:background}, we provide a short review of relevant results from the equilibrium theory of wetting. In Section \ref{sec:model} we define precisely the hydrodynamic model studied here, as well as its numerical implementation. Section \ref{sec:results} then presents the surface phase diagram and characterises the wetting states and their transition, as well as introducing the nonequilibrium effects evident in active wetting. Finally, Section \ref{sec:discussion} summarises our conclusions, discussing the results presented here and in \cite{Grodzinski2026a}, and directions for continued research.

\section{Background -- equilibrium wetting}\label{sec:background}
This section reviews the \textit{equilibrium} theory of wetting, to provide context to our work and points of comparison. We focus mainly on results derived from the Cahn theory of wetting \cite{Cahn1977}, as this (mean-field, noise-free) model is most analogous to our exact hydrodynamic description.

Wetting describes the equilibrium behaviour of a fluid with coexisting liquid and vapour phases when it comes into contact with an attractive solid wall~\cite{Evans2019, Cahn1977, DeGennes1985, Dietrich1988, Bonn2009}. 
The liquid typically forms droplets on the wall, characterised by a contact angle $\theta_w$ (see Figure \ref{fig:wetting_schematic}(a)).  
This contact angle can be calculated from the interfacial tensions between the liquid, vapour, and solid phases, and is given by Young's equation \cite{DeGennes1985}
\begin{equation}\label{eq:young}
    \cos(\theta_w) = \frac{\gamma_{sv} - \gamma_{sl}}{\gamma_{lv}},
\end{equation}
where $\gamma_{lv}$, $\gamma_{sv}$, and $\gamma_{sl}$ denote the liquid–vapour, solid–vapour, and solid–liquid interfacial tensions respectively. Young’s equation can be derived either mechanically, by considering the balance of forces at the contact line \cite{DeGennes1985}, or using statistical physics, via the variational minimisation of a free energy \cite{DeGennes2003}. These perspectives are equivalent because the system is at equilibrium.

When $\theta_w \in (0, \pi)$, the liquid forms a droplet on the wall; this is called partial wetting.
A wetting transition occurs when the contact angle $\theta_w$ vanishes, and the liquid spreads across the wall to form a macroscopic film; this state with $\theta_w=0$ is called full wetting.  A drying transition occurs as $\theta_w$ approaches $\pi$ and the droplet detaches from the wall.

\begin{figure}[t]
\centering
\begin{tikzpicture}[scale=1.5]

\fill[gray!20] (-2,-0.6) rectangle (2,0);
\draw[thick] (-2,0) -- (2,0);
\node at (0,-0.35) {Solid surface};

\draw[thick, fill=blue!25, domain=30:150] plot ({cos(\x)}, {sin(\x) - 0.5});
\node[blue!60!black] at (-0.3,0.7) {Liquid};

\node at (0.5,0.18) {\small $\theta_w$};

\node at (1.5,1) {Vapour};

\draw[->, thick, red] (0.866,0) -- (0.866-0.4,0) node[below left] {\small $\gamma_{sl}$};
\draw[->, thick, red] (0.866,0) -- (0.866+0.7,0) node[below right] {\small $\gamma_{sv}$};
\draw[->, thick, red] (0.866,0) -- (0.866-0.4,0.6) node[right] {\small $\gamma_{lv}$};

\draw[thick] (0.4,0) ++(-0.25:0.25) arc[start angle=180, end angle=130, radius=0.25];

\node at (-2.1,1.2) {{(a)}};

\end{tikzpicture}
\vspace{0.8cm}

\begin{tikzpicture}[scale=1.1]

\node at (-2,1.8) {\text{(b.1) Fully wet}};
\node at (0.9,1.8) {\text{(b.2) Partially wet}};
\node at (3.4,1.8) {\text{(b.3) Dry}};

\node at (-1.9,-1.8) {$F = 2 \gamma_{sl} + 2 \gamma_{lv}$};
\node at (0.9,-1.8) {$F = \gamma_{sl} + \gamma_{lv} + \gamma_{sv}$};
\node at (3.6,-1.8) {$F = 2 \gamma_{sv} + 2 \gamma_{lv}$};

\fill[gray!20] (-3.0,-1.5) rectangle (-2.7,1.5);
\fill[gray!20] (-1.0,-1.5) rectangle (-0.7,1.5);
\fill[blue!25] (-2.7,-1.5) rectangle (-2.3,1.5);
\fill[blue!25] (-1.4,-1.5) rectangle (-1.0,1.5);
\draw[thin] (-2.7,-1.5) -- (-2.7,1.5);
\draw[thin] (-1.0,-1.5) -- (-1.0,1.5);
\node[rotate=90] at (-2.85,0) {\footnotesize Solid Wall};
\node[rotate=90] at (-0.85,0) {\footnotesize Solid Wall};
\node[blue!60!black, rotate=90] at (-2.5,0) {\footnotesize Liquid};
\node[blue!60!black, rotate=90] at (-1.2,0) {\footnotesize Liquid};
\node[rotate=90] at (-1.85,0) {\footnotesize Vapour};

\fill[gray!20] (-0.3,-1.5) rectangle (0.0,1.5);
\fill[gray!20] (1.7,-1.5) rectangle (2.0,1.5);
\fill[blue!25] (0.0,-1.5) rectangle (0.8,1.5);
\draw[thin] (0.0,-1.5) -- (0.0,1.5);
\draw[thin] (1.7,-1.5) -- (1.7,1.5);

\fill[gray!20] (2.4,-1.5) rectangle (2.7,1.5);
\fill[gray!20] (4.4,-1.5) rectangle (4.7,1.5);
\fill[blue!25] (3.1,-1.5) rectangle (4.0,1.5);
\draw[thin] (2.7,-1.5) -- (2.7,1.5);
\draw[thin] (4.4,-1.5) -- (4.4,1.5);

\draw[thick] (-3,-1.5) -- (-0.7,-1.5) -- (-0.7,1.5) -- (-3,1.5) -- cycle;
\draw[thick] (-0.3,-1.5) -- (2,-1.5) -- (2,1.5) -- (-0.3,1.5) -- cycle;
\draw[thick] (2.4,-1.5) -- (4.7,-1.5) -- (4.7,1.5) -- (2.4,1.5) -- cycle;

\end{tikzpicture}
\caption{(a) Diagram of equilibrium wetting on a solid surface. A liquid droplet rests on the solid in coexistence with the vapour phase, forming a contact angle $\theta_w$. The interfacial tensions $\gamma_{sl}$, $\gamma_{sv}$, and $\gamma_{lv}$ correspond to solid–liquid, solid–vapour, and liquid–vapour interfaces, respectively. (b)~Schematic of wetting states in a slit geometry with a fixed mass of fluid. (b.1)~Fully wet: macroscopic liquid films coat both walls. (b.2)~Partially wet: a macroscopic liquid film forms on one side. (b.3)~Dry (Unpinned): no bulk liquid wets the walls.}
\label{fig:wetting_schematic}
\end{figure}

Wetting can alternatively be studied in the context of a fluid in a slit geometry at constant total mass (i.e. a canonical ensemble) \cite{Leeuwen1989, Nijmeijer1990, Bucior2009, Swain2000, Valencia2001, Sikkenk1987},
where the fluid is confined between two parallel planar walls, separated by a macroscopic distance $L$.  The system is assumed to be translationally symmetric in the direction(s) parallel to the walls;
the total fluid mass, and other control parameters, are chosen such that both a bulk liquid and bulk vapour phase coexist. This allows for three possible wetting states (see Figure \ref{fig:wetting_schematic}(b), and \cite{Evans2019, Leeuwen1989, Nijmeijer1990} for a more detailed discussion):
\begin{enumerate}[(i)]
  \item \textit{Fully wet}: The fluid forms a macroscopic liquid layer on both walls with a vapour bulk between them, such that the total energy per unit (transverse) area is $2 \gamma_{sl} + 2 \gamma_{lv}$.
  \item \textit{Partially wet}: The fluid forms a macroscopic liquid layer on one wall while the other remains dry\footnote{{Although this ``dry'' wall has no liquid bulk attached, a thin adsorbed film remains present within the partially wet regime. This leads to the relevant ``moist'' surface tension $\gamma_{sv}$ \cite{DeGennes1985}.}}, such that the total energy per unit area is $\gamma_{sl} + \gamma_{lv} + \gamma_{sv}$.
  \item \textit{Dry/Unpinned}: Both walls are dry, and there is an unpinned liquid bulk between the walls such that the total energy per unit area is $2 \gamma_{sv} + 2 \gamma_{lv}$.
\end{enumerate}

Comparing the energies, the transition from partial to full wetting in the slit geometry occurs at $\gamma_{sv} = \gamma_{sl} + \gamma_{lv}$ which corresponds to $\theta_w=0$, as for the droplet. The wetting transitions in the slit and droplet geometry therefore correspond to the same physical transition. The same applies to the drying transition {between states (ii) and (iii)}. We emphasise that a wet wall is covered by a macroscopic liquid layer of thickness $\zeta=\mathcal{O}(L)$. For attractive walls,  even a ``dry'' wall is typically coated by a microscopic film of fluid, $\zeta=\mathcal{O}(1)$ \cite{DeGennes1985}.

Wetting transitions occur as some control parameter is varied (for example, the strength of an attractive interaction between fluid and wall); we denote this quantity here by $\epsilon$.  
It is useful to define a spreading factor $S=\gamma_{sv} - \gamma_{sl} - \gamma_{lv}$ ~\cite{DeGennes1985}, which is equal to
\begin{equation}
   S = \gamma_{lv} [ \cos(\theta_w) - 1] ,
\end{equation}
in the partially wet state, with $S<0$. A wetting transition means that $S\to0^-$ as a control parameter $\epsilon$ tends to a critical value $\epsilon^*$.  These transitions come in two types:
first-order wetting involves $S\to0$ with a finite slope, $dS/d\epsilon\neq 0$; in this case the fluid film thickness $\zeta$ (on the ``dry'' side) approaches a finite limit as $S\to0$ and then jumps to a macroscopic value $\zeta=O(L)$ only in the fully-wet state. Critical\footnote{{Note that these wetting transitions are critical as they involve divergence of the film width as a surface parameter is varied, at two-phase coexistence. This is distinct from the complete wetting transition which occurs approaching coexistence from a single phase, e.g. \cite{Neta2021}.}} transitions have $S\to0$ and $dS/d\epsilon\to 0$ together; this signals a continuous divergence of the thickness $\zeta$, see below.

Cahn \cite{Cahn1977, DeGennes1985} showed that the equilibrium wetting transition of a fluid on a planar wall, or in the slit geometry, can be described at hydrodynamic level by a one-dimensional field theory for the fluid density $\rho$, with free energy
\begin{equation}\label{F_eq}
    F[\rho] = \int_0^\infty \frac{\kappa}{2} (\nabla \rho)^2 + W(\rho) + f_w(\rho) \delta(x) \dd{x},
\end{equation}
where $W(\rho)$ is a double-welled potential describing bulk phase separation. The free energy due to interaction with the wall (at $x=0$) is contained in the function $f_w(\rho)$, which can be expanded in a Taylor series as
\begin{equation}\label{gamma_c}
  f_w(\rho) = c_0 - c_1 \rho + \frac{1}{2} c_2 \rho^2 + \mathcal{O}(\rho^3).
\end{equation}
Here $c_1$ roughly corresponds to the attractive interactions between the wall and the fluid, and $c_2$ to the effect of exclusion of fluid by the wall, resulting in the loss of attractive fluid-fluid interactions (see \cite{DeGennes1985}). Taking the control parameter $\epsilon = c_1$ (so that the wetting transition occurs on varying the attraction strength),  $\epsilon<\epsilon^*$ corresponds to partial wetting (weaker attraction) and $\epsilon>\epsilon^*$ to full wetting.

The wetting transition in a slit geometry, where impenetrable walls are placed at $x = 0,L$, can then be characterised using a global asymmetry
order parameter~\cite{Turci2021}
\begin{equation}\label{eq:asy}
  \mathcal{A}[\rho] = \frac{\int_0^L \abs{\rho(x) - \rho(L-x)} \dd{x} }{2L(\bar{\rho} - \rho_v)}.
\end{equation}
Here $\mathcal{A}=0$ characterises full wetting, $\mathcal{A}>0$ indicates partial wetting, and the normalisation is such that $\mathcal{A} \in [0,1]$.

Minimising the free energy \eqref{F_eq} results in an equilibrium theory of wetting with mean-field character (because the hydrodynamic approach neglects density fluctuations).  This theory is summarised in Appendix~\ref{app:wetting}; both critical and first-order wetting are possible in the Cahn theory, with critical wetting observed if $c_2$ is sufficiently large. Physically, {this} corresponds to the scenario where the effect of the wall in preventing attractive interactions between nearby fluid particles is large.

In later sections, we will compare the active wetting transition to the equilibrium critical wetting transition.
To facilitate this discussion, we identify here the main distinguishing features of critical wetting at equilibrium (as opposed to first-order wetting), which can be derived from the Cahn theory in a slit with constant total mass. First, the film thickness on the dry wall diverges continuously at the wetting transition, as
\begin{equation}\label{eq:log_div}
  \zeta \sim \log\frac{1}{\epsilon^*-\epsilon} , \qquad \epsilon<\epsilon^* \; .
\end{equation}
Second, the asymmetry parameter vanishes continuously as full wetting is approached (exhibiting a pitchfork bifurcation with a characteristic exponent of $1/2$):
\begin{equation}
  \mathcal{A} = \begin{cases}
    \alpha (\epsilon^* - \epsilon)^{1/2}, & \; \epsilon < \epsilon^*, \\
    0, & \; \epsilon \geq \epsilon^*, 
  \end{cases}
  \label{eq:Aeq-sqrt}
\end{equation}
where $\alpha$ is a proportionality constant.
Third, there are no metastable states so one does not expect hysteresis as $\epsilon$ passes through $\epsilon^*$.

The theory of wetting has been extended to understand numerous surface phenomena beyond those described above. See \cite{Bonn2009} for a modern review. For example, away from liquid-vapour coexistence in a grand canonical ensemble, \textit{prewetting} {or \textit{complete wetting}} transitions may still be observed (see {\cite{Evans1983, Nakanishi1982, Dietrich1988, Sullivan1986, Binder1988}}), or slits may fill entirely with liquid in a phenomenon known as \textit{capillary condensation} \cite{Evans1985, Malijevsky2021}. These effects are not relevant for our model as we work at liquid-vapour coexistence. {We include in Appendix \ref{app:wetting_categories} a discussion of the distinct surface phenomena which are possible in equilibrium, and in Appendix \ref{app:slabs_droplets} a discussion of the different partially wet morphologies which are possible in a slit geometry with 2 spatial dimensions.} Noise, dimensionality, and the nature/range of particle-particle/particle-wall interactions may all subtly change the order of wetting and drying transitions -- see, for example, \cite{Wu2016, Evans2019, Binder2003}.

\section{Model and methods}\label{sec:model}

\subsection{Active lattice gas model}\label{subsec:active_dynamics}
\subsubsection{Exact hydrodynamic description}
We study the active lattice gas introduced in~\cite{Mason2023a}, which involves particles that move on an $N \times N$ square lattice, with at most one particle per site, modified by the presence of an external potential $V$.  We embed the lattice into a domain of size $L\times L$, so that the lattice spacing is $L/N$.  The hydrodynamic limit is $N\to\infty$ at fixed $L$.  Each particle $i$ is endowed with a position $\vb{x}_i$ and an orientation vector $\vb{e}(\theta_i) = (\cos(\theta_i), \sin(\theta_i))^T$.  It hops from $\vb{x}_i$ to a vacant neighbouring site $\vb{x}_i+\vb{u}$ with rate 
\begin{equation}
w_i(\bm{u}) = N^2\left\{ D_E + \frac{v_0}{2} \vb{u} \cdot \vb{e}(\theta_i) + \frac{V(\vb{x}_i) - V(\vb{x}_i+\vb{u})}{2} \right\}.
\end{equation} 
(Note that $|\bm{u}|=L/N$ is small in the hydrodynamic limit.)
Each angle $\theta_i$ undergoes Brownian motion on the unit circle with orientational diffusion constant $D_O$, and spatial diffusion of particle positions $\vb{x}_i$ occurs with diffusion constant $D_E$ (as above). The only interactions between particles are excluded volume interactions.
On taking the hydrodynamic limit,  it is important that particles' movement rates are chosen to have a non-trivial $N$-dependence, so that both diffusive and self-propelled contributions remain comparable as $N\to\infty$.  This feature enables the exact calculation; it involves a limit of fast diffusion that suppresses fluctuations so that the resulting system is of mean-field type, similar to the Cahn theory. We note that, due to this suppression of noise-induced fluctuations  in the hydrodynamic limit, the bubbly phases seen in some active systems \cite{Cates2025, Tjhung2018, Yao2025} are not possible here, even though bubbles may remain present in the limit of a large system size for some parameter ranges \cite{Shi2020} (see also Sec.~\ref{subsec:outlook}). Our dynamics also exclude inertial effects; while inertia is known to influence wetting behaviour in active systems \cite{Caprini2024}, this is not considered here.

The system is described in the hydrodynamic limit by a probability density $f(\vb{x}, \theta, t)$ for particles with spatial position $\vb{x}$ and orientation $\theta$ at time $t$.
The spatial density $\rho$ and polarisation $\vb{p}$ of the particles are
\begin{align}
  \rho(\vb{x}, t) &= \int_0^{2\pi} f(\vb{x}, \theta, t) \dd{\theta}, \\
  \vb{p}(\vb{x}, t) &= \int_0^{2\pi} f(\vb{x}, \theta, t) \ {\vb{e}}(\theta) \dd{\theta}.
\end{align}
(Note that $\rho(\vb{x},t) \in [0, 1]$; this quantity represents the average occupancy of sites in the vicinity of position $\vb{x}$.  The number density of particles is diverging proportional to $N^2$ in the hydrodynamic limit.)
The deterministic evolution of $f$ is given by \cite{Mason2023a}
\begin{multline}\label{eq:dynamics_from_S}
   \partial_t f = \div \int  \sigma_f(\vb{x},\theta,\theta') \qty[ D_E \grad \frac{\delta \mathcal{S}}{\delta f(\vb{x},\theta')} + v_0 \vb{e}_{\theta'} ] d\theta' \\ +  D_O \partial_\theta \qty[  f(\vb{x},\theta) \partial_\theta \frac{\delta \mathcal{S}}{\delta f(\vb{x},\theta)} ] ,
\end{multline} 
where the free energy is
\begin{multline}\label{eq:S}
  \mathcal{S}[f] = \int \dd{\vb{x}} \{ [1-\rho(\vb{x})] \log[1-\rho(\vb{x})] + \rho(\vb{x})V(x)\}  \\ +  \int \dd{\vb{x}} \dd{\theta} f(\vb{x}, \theta) \log(2 \pi f(\vb{x}, \theta) )
\end{multline} 
and the quantity
\begin{multline} \sigma_f(\vb{x},\theta,\theta') \equiv s(\rho(\vb{x})) f(\vb{x},\theta) f(\vb{x},\theta') \\ + d_s(\rho(\vb{x})) f(\vb{x},\theta) \delta(\theta - \theta') \end{multline}
plays the role of a mobility. We define $d_s(\rho)$ as the diffusion constant in a symmetric exclusion process at density $\rho$ (see below), and use the shorthand 
\begin{equation}
    s(\rho) := \frac{1 - d_s(\rho) - \rho}{\rho}.
\end{equation}
This formulation of the equation of motion allows the terms involving $\delta \mathcal{S}/\delta f$ to be identified with equilibrium-like processes, which act to minimise $\mathcal{S}$.  By contrast, the term proportional to $v_0$ is the nonequilibrium self-propulsion force.

The potential $V$ enters the dynamics as an extra equilibrium-like force, adding an additional contribution to the free energy \eqref{eq:S} compared to an otherwise equivalent expression in \cite{Mason2023a}. The added potential results in an extra contribution of $V(\vb{x})$ to $\frac{\delta \mathcal{S}}{\delta f(\vb{x},\theta')}$; the additional term is independent of $\theta'$, which reflects that the force from the barrier does not depend on the particle orientation.  

Writing \eqref{eq:dynamics_from_S} explicitly, the deterministic dynamics of $f$ is given by:
\begin{equation}\label{eq:ABP_evolution}
  \pdv{f}{t} = - D_E \div \vb{J}_\text{diff} - v_0 \div \vb{J}_\text{activity} + D_O \partial_\theta^2 f,
\end{equation}
where
\begin{align}
  &\vb{J}_\text{diff} = -d_s(\rho) \grad f - f \mathcal{D}(\rho) \grad \rho - f(1 - \rho) \grad V, \label{eq:J_diff}\\ 
  &\vb{J}_\text{activity} = (\mathcal{D}(\rho) - 1) f \vb{p} + \vb{e}_\theta d_s(\rho) f. \label{eq:J_act}
\end{align}
Here we take $ \mathcal{D}(\rho) := [1 - d_s(\rho)]/\rho$ for ease of writing. The fact that $\int_0^{2\pi} \sigma_f(\rho,\theta,\theta') d\theta' = f(\theta)$ explains why $V$ appears in such a simple way in Eq.~\eqref{eq:J_diff}. It was shown in~\cite{Mason2023b} that the self-diffusion coefficient can be approximated as:
\begin{equation}\label{eq:d_s_rho}
d_s(\rho) = (1-\rho) \left( 1 - \alpha \rho + \frac{\alpha(2 \alpha - 1)}{2 \alpha + 1} \rho^2 \right),
\end{equation}
where $\alpha = \frac{\pi}{2} - 1$. 
This formula for $d_s$ is exact in both the low- and high-density limits and gives accurate approximations for all densities. We will use it below whenever an explicit form for $d_s(\rho)$ is required.

We identify two length scales associated with the particle dynamics:
\begin{equation}
  l_p = \frac{v_0}{D_O}, \qquad l_D = \sqrt{\frac{D_E}{D_O}},
\end{equation}
where
$l_p$ corresponds to the typical distance a particle moves due to self-propulsion before changing direction, and $l_D$ corresponds to the typical distance a particle moves due to diffusion in the reorientation time. 

It is convenient to rescale time as $t \rightarrow D_O t$, and space as $x \rightarrow \frac{x}{l_D}$, such that both are made dimensionless.  The hydrodynamic equation \eqref{eq:ABP_evolution} 
becomes
\begin{equation}\label{eq:dynamics_dimensionless}
  \pdv{f}{t} = - \div \vb{J}_\text{diff} - \text{Pe}\, \div \vb{J}_\text{activity} + \partial_\theta^2 f,
\end{equation}
where 
\begin{equation}
\text{Pe}  = \frac{l_p}{l_D} = \frac{v_0}{\sqrt{D_E D_O}}
\end{equation}
is the Péclet number, describing the dimensionless strength of self-propulsion.  We work with the rescaled space and time dynamics \eqref{eq:dynamics_dimensionless} throughout the following.

\begin{figure}[t]
  \centering
  \includegraphics[width=1.0\linewidth]{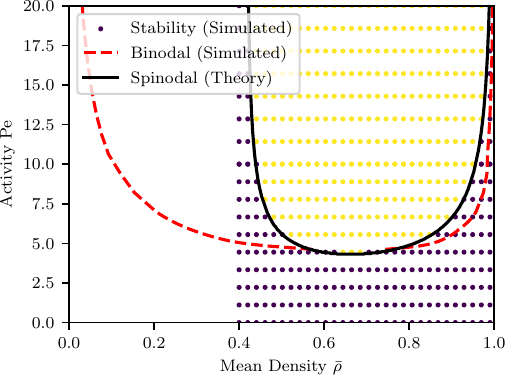}
  \caption{Phase diagram of the active lattice gas.  The binodal (coexistence curve) is obtained by solving \eqref{eq:ABP_evolution}-\eqref{eq:J_act} numerically, and characterising the phase-separated steady-states.  The spinodal is obtained from~\cite{Mason2023a}, it is compared with numerical stability results for \eqref{eq:ABP_evolution}-\eqref{eq:J_act} starting from weakly perturbed homogeneous states (coloured points, no data shown for $\bar{\rho} < 0.4$).
  The critical point where the spinodal and binodal curves intersect is ($\bar{\rho}_\text{crit}, \text{Pe}_\text{crit}) = (0.659, 4.33)$. As is usual for active phase separation \cite{Cates2025}, these binodals (dashed curves) do not depend on the value $\bar{\rho}$ so long as it lies between them. See Appendix \ref{sec:binodal_spinodal_convergence} for further details. }
  \label{fig:binodal_spinodal}
\end{figure}

The active lattice gas exhibits motility-induced phase separation (MIPS), arising from the interplay between the self-propulsion of the particles and the (purely repulsive) excluded volume interactions between them \cite{Cates2015, Cates2013, Mason2023a}.
The spinodal curve was computed in \cite{Mason2023a}. The binodal can be computed by simulating the hydrodynamic equation~\eqref{eq:ABP_evolution}-\eqref{eq:J_act} -- see Appendix \ref{sec:binodal_spinodal_convergence}.  The resulting phase diagram is shown in Figure \ref{fig:binodal_spinodal}. Generically, we expect the width of liquid-vapour interfaces to scale as $\xi \sim l_D f_\xi(\text{Pe})$, and diverge as $\text{Pe} \to \text{Pe}^*(\bar{\rho})$ \cite{Mason2023a}.

\subsubsection{System geometry and wall/barrier potential}\label{sec:choice_permeable_barrier}

Following~\cite{Turci2021, Das2020}, we study a two-dimensional system with periodic boundary conditions, and a penetrable wall (or ``barrier'') located at $x=0$.  In the limit of large barrier height, this becomes the slit geometry of Figure ~\ref{fig:wetting_schematic}(b).  It is known that active particles generically adsorb to hard walls \cite{Das2018, Elgeti2013, Sepulveda2017, Sepulveda2018}, leading generically to full {(complete)} wetting {in the context of 2-phase coexistence} \cite{Zhao2026, Turci2021, Das2020, Neta2021, Turci2024, PerezBastias2025}. Partial wetting states have been observed in active systems, either with soft walls \cite{Zhao2026}, or using penetrable barriers \cite{Turci2021, Das2020}. We choose a smooth, penetrable barrier in order to access a transition between full- and partial-wetting states. Periodic boundary conditions are used to avoid introducing any additional confining boundaries. Experimental realisations of this geometry are discussed in Section \ref{subsec:outlook}. We emphasise that our study is at fixed total fluid mass and at liquid-vapour coexistence (in contrast to, e.g. \cite{Neta2021}), and so the phenomena of prewetting and capillary condensation \cite{Bonn2009} are not relevant here. 

\begin{figure}[t]
  \centering
  \includegraphics[width=1.0\linewidth]{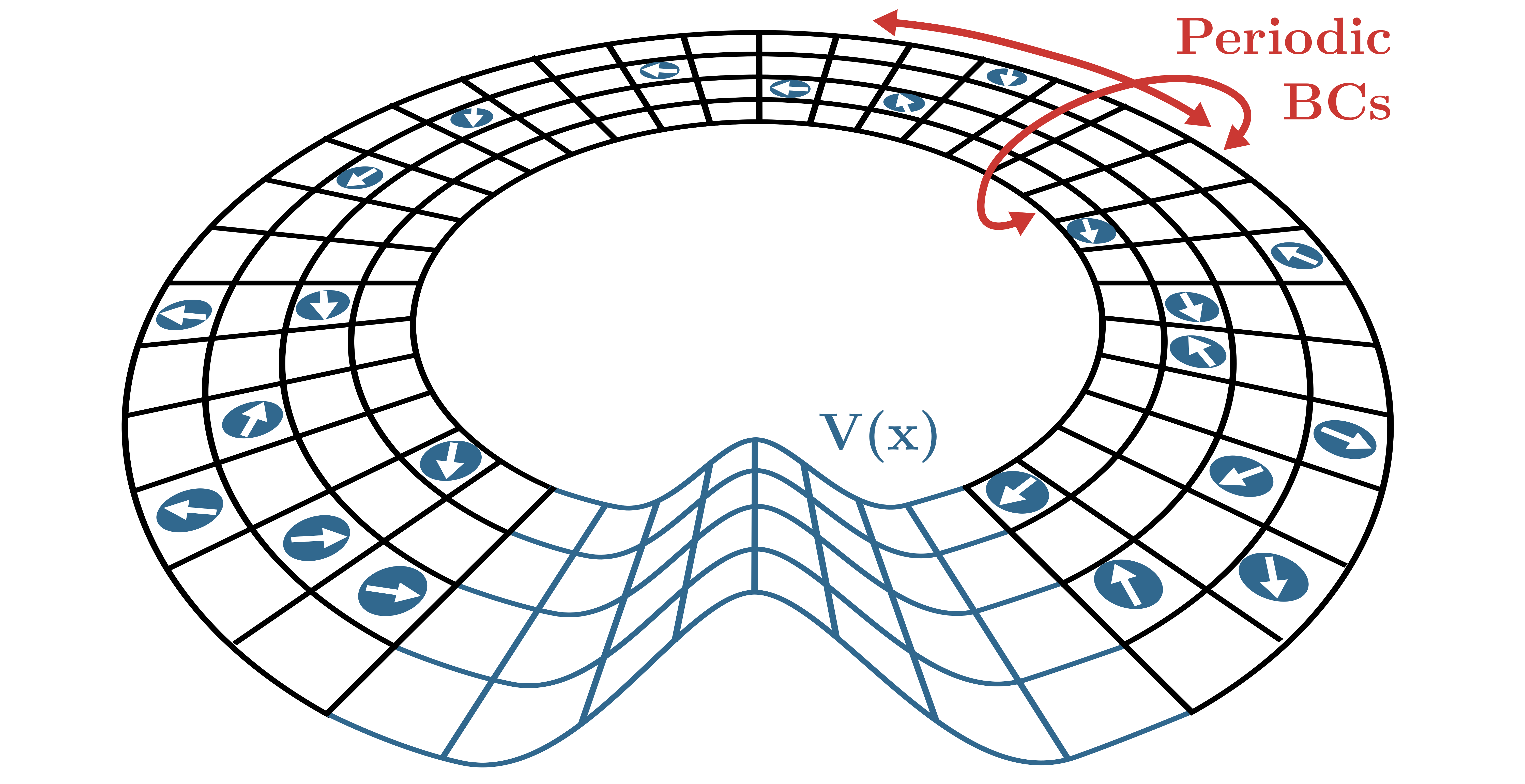}
  \caption{Schematic illustration of the system geometry. The system occupies a two-dimensional periodic domain, with a smooth permeable barrier centred at $x=0$. The barrier is represented by an external potential $V(x)$ of finite width $w_b$ and height $\epsilon$, which locally repels particles while allowing finite probability of crossing. Active particles move with self-propulsion $v_0$ along the $\vb{e}_\theta$-direction and interact only via excluded volume. The orientation-dependent probability density $f(\vb{x}, \theta, t)$ is translationally invariant in the transverse direction(s), so the wetting behaviour is fully characterised by the one-dimensional profile shown here.}
  \label{fig:cartoon}
\end{figure}

Since the governing PDEs are deterministic and the initial condition is chosen to be translationally invariant along the direction parallel to the walls, the probability density $f(\vb{x},\theta,t)$ only depends on the component $x$ of the position $\vb{x}$ perpendicular to the barrier.  Hence, we consider $1d$ density profiles $f(x, \theta, t)$ in the following, although we emphasise that the underlying lattice model is $2d$ and can easily be generalised to higher dimensions \cite{Mason2023a, KourbaneHoussene2018}, {see Appendix \ref{app:slabs_droplets}}. (Indeed, it is not possible to have orientational diffusion at fixed propulsion speed $v_0$ in one spatial dimension.) See Section \ref{sec:background} above for a discussion of equilibrium wetting with one non-translationally-invariant direction. We choose a system with domain size $L$ in the $x$-direction.

The barrier potential is taken as in~\cite{Turci2021}:
\begin{equation}\label{eq:potential}
  V(x) = \begin{cases}
    0 & \text{if } \abs{x} \geq w_b, \\
    \epsilon \qty[\cos(\frac{\pi x}{w_b}) + 1] & \text{if } \abs{x} < w_b,
  \end{cases}
\end{equation}
so the strength of the barrier potential is a dimensionless parameter $\epsilon$; its half-width $w_b$ is fixed throughout as $w_b=l_D/2$.  The final (dimensionless) control parameter, alongside the barrier strength $\epsilon$ and the activity level $\text{{Pe}}$, is the (conserved) total density
\begin{equation}
\bar{\rho} \equiv \frac{1}{L} \int_0^L \rho(x) \dd{x}.
\end{equation}

In summary, our system has four control parameters. Two intensive quantities, the Péclet number $\text{Pe}$ and mean density $\bar{\rho}$, control particle behaviour. The geometry is controlled by the barrier height $\epsilon$ and the (reduced, dimensionless) system size $\ell_s = L/l_D$. The latter is generally chosen such that the slit width is much larger than the scale of liquid-vapour interfaces (although its exact value depends on the context -- see Sections \ref{subsec:numerical_implementation} and \ref{sec:results}), which is the thermodynamic limit.

\subsection{Details of numerical implementation}\label{subsec:numerical_implementation}
To solve the hydrodynamic equation~\eqref{eq:ABP_evolution}, we discretise the orientation-dependent
density $f(x,\theta,t)$ onto an $N_x \times N_\theta$ grid, and integrate
\eqref{eq:ABP_evolution}-\eqref{eq:J_act} using an explicit first–order upwind
update scheme. Further details are given in Appendix \ref{app:timestepping}, along with computational parameters. Writing the evolution equation as
\begin{equation}
    \partial_t f = -\partial_x \!\left[M_x U_x\right]
               -\partial_\theta \!\left[M_\theta U_\theta\right]
\end{equation}
allows separate treatment of the spatial and orientational dynamics; we choose mobilities $M_x=f\,d_{\rm s}(\rho)$ and $M_\theta=f$ for numerical stability (in particular, the presence of $\log(f)$ in the resulting velocity fields strongly protects the dynamics against negative densities).
A Courant–Friedrichs–Lewy (CFL) condition controls the adaptive time step and protects against advective instabilities.  
We verified that this method conserves the total density to high accuracy, preserves the
parity symmetry of steady-states, and reproduces the expected linear
stability properties of the homogeneous active fluid (i.e. the spinodal matches the theoretical prediction given in \cite{Mason2023a}).

\paragraph*{Determination of the binodal and spinodal lines.}
For the bulk (barrier-free) system, the spinodal is obtained from the linear
stability analysis of~\cite{Mason2023a}. The onset of instability is numerically verified by perturbing a homogeneous state with the maximally unstable\footnote{That is the eigenfunction whose eigenvalue in the linearised dynamics is largest, even if this is negative (which we expect to be stable).} eigenfunction, and measuring the growth rate of this instability. The numerical results are in good agreement with the theoretical spinodal (see Figure \ref{fig:binodal_spinodal}).

The binodal is computed numerically (a theoretical prediction is not currently available); starting from a weakly perturbed
homogeneous state inside the spinodal, the system is evolved until a
liquid–vapour interface forms.  
This profile is then embedded in a larger system with reflecting boundaries,
yielding well-converged estimates of the coexisting densities $\rho_{v}$ and
$\rho_{l}$.  
These values depend only on ${\rm Pe}$, consistent with
standard MIPS phenomenology. These results are presented in Figure \ref{fig:binodal_spinodal}.

\paragraph*{Classification of wetting states.}
Because direct simulation to asymptotically long times is computationally
costly, we classify steady-states using measurements taken at a finite final
time $t_f$.  
States are first distinguished as homogeneous or phase-separated via the
variance of the density away from $\bar\rho$, excluding the barrier region.
For phase-separated states, wetting of the barrier is detected from the
behaviour of $\rho(x)$ near $x=0$; finally, full versus partial wetting is
identified using the asymmetry order parameter $\cal A$.
The wetting transition line (distinct from the classification of individual states) is obtained by testing the linear stability of the
fully wet state: whenever small asymmetries about the barrier grow in time, the system lies
in the partially wet regime.
Comparison with longer simulations confirms that this procedure reliably
identifies the long-time behaviour except within a narrow neighbourhood of the
transition lines, and that these classifications are unchanged when the simulations are run with a larger system size or using different values of the computational parameters. Further details can be found in Appendix \ref{app:timestepping}.

\begin{figure*}[t]
    \centering
      \includegraphics[width=\linewidth]{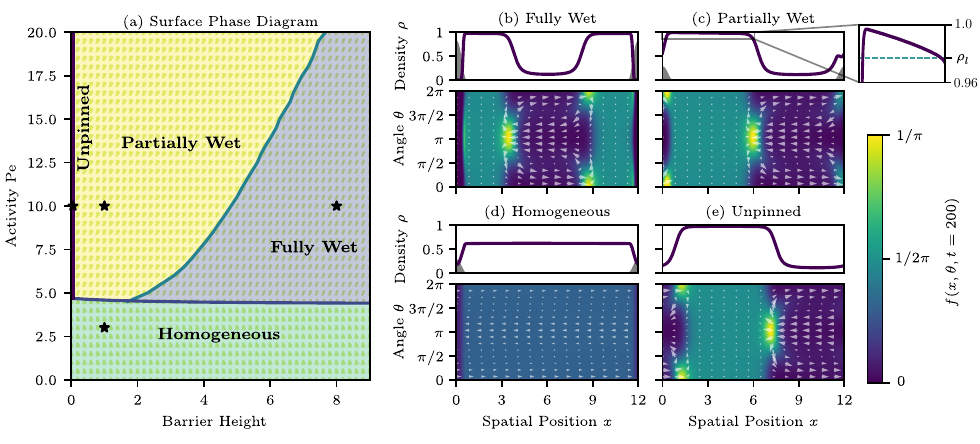}
      \caption{Wetting of a barrier by the active lattice gas.
      (a) Surface phase diagram in the $(\epsilon,\text{Pe})$ plane. Each point represents a numerical solution (simulation) of the hydrodynamic equation, whose final state is classified with the routine described in Section \ref{subsec:numerical_implementation}, which also explains the estimation of the phase boundaries. The blue line is the onset of MIPS; the green line is the wetting transition; the dark purple line likewise represents unpinning of the liquid at $\epsilon = 0$. The structure of this surface phase diagram does not depend on $\bar{\rho}$, as long as it remains inside the spinodal. See Section \ref{subsec:numerical_implementation} for details of numerics.
      (b-e) Snapshots of the steady-state densities $f(x, \theta)$ and $\rho(x)$ for the starred points shown in the phase diagram, which illustrate the four phases. The upper plots show the density profile $\rho(x)$, and the potential $V(x)$; the lower plots show $f(x, \theta)$ (colour) and ${\cal J}(x, \theta)$ (arrows -- scaled proportionally to $\mathcal{J}^{1/4}$). The magnified section of (c) shows a linear density gradient throughout the liquid bulk, in the partially-wet state; it shows $x\in [0,6]$.}
      \label{fig:phase_behaviour}
  \end{figure*}

\section{Results}\label{sec:results}

We access the wetting transition by varying the self-propulsion $\text{Pe}$ and the barrier height $\epsilon$.  We fix the total density at its critical value $\bar\rho^*=0.659$, identified via linear stability analysis (see Section \ref{subsec:numerical_implementation} and \cite{Mason2023a}); we mostly take $\text{Pe}$ above its critical value, in which case results depend weakly on $\bar\rho$, provided it remains inside the spinodal (recall Fig.~\ref{fig:binodal_spinodal}). Repeating the analysis at different values of $\bar{\rho}$ inside the spinodal leads only to small movements of the transition lines, instead of changes to the structure of the phase diagram. 
We choose different values of the reduced system size $\ell_s$ according to the context, but these are always large enough that we recover bulk phase separation, and so that the interfacial width is much smaller than the slit width. 

\begin{figure*}[t]
    \centering
    \includegraphics[width = \linewidth]{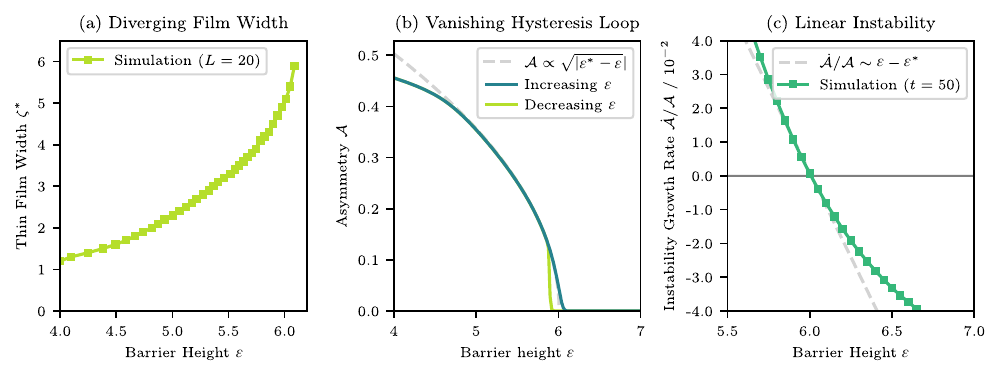}
    \caption{The wetting transition between fully- and partially-wet states is critical.
    All plots are at $\text{Pe} = 10$.
    (a) The width of the liquid film on the ``dry'' wall, estimated by measuring the distance between the inflection points of $\rho$ at the liquid-boundary and liquid-vapour interfaces. This increases as the transition is approached.  Results are shown for a system of size $\ell_s=20$, so maximum width of (symmetric) liquid layers is approximately $6.3$ (using the lever rule). For an infinite system we expect $\zeta$ to diverge at $\epsilon = \epsilon^*$.
    (b) Asymmetry $\mathcal{A}$ as a function of $\epsilon$, scanning both up (green) and down (orange), to check for hysteresis.  The total scan time is $T_\text{tot} = 2 \times 10^5$; we use a system size $\ell_s = 5$, smaller than in (a), such that the scan is quasi-static and the behaviour of $\mathcal{A}$ is not too sharp (see Appendix \ref{app:wetting}).
    The dashed grey line is a fit to \eqref{eq:Aeq-sqrt}. We estimate $\epsilon^* \approx 6.01$. The hysteresis loop is observed to shrink as the scan time is increased. (c) Linear stability of the fully-wet state with the same parameters.  We plot the exponential growth rate of the asymmetry $\dot{\mathcal{A}}/\mathcal{A}$ as a function of barrier height $\epsilon$.  Positive values correspond to partial wetting (the fully-wet state is unstable), and we estimate $\epsilon^* \simeq 6.01$ from a linear fit.  The growth rate is measured at $t=50$, when asymmetry growth is approximately linear and $\mathcal{A}$ remains small.}
    \label{fig:wetting_transition}
\end{figure*}
\subsection{Full and partial wetting}\label{subsec:active_wetting_states}

Figure \ref{fig:phase_behaviour}(a) shows the surface phase diagram, for a range of $\epsilon, \text{Pe}$.
To obtain this, we initialise the system in a homogeneous state
\begin{equation}
    f(x, \theta, 0) = \frac{\bar{\rho}_\text{crit}}{2\pi}+ \delta f,
\end{equation}
where $\delta f \sim 10^{-4}$ is a small perturbation, uncorrelated in space. The system is then evolved to a final time $t_{\rm f} = 200$, and states/transition lines are classified according to the routine described in Section \ref{subsec:numerical_implementation}, based on an extrapolation to the large time limit. We identify four distinct steady-states, analogous to the states found in equilibrium wetting. These are labelled fully-wet, partially-wet, unpinned, and homogeneous. They are illustrated in Figures \ref{fig:phase_behaviour}(b–e).  

The fully-wet state is symmetric, with liquid layers of equal width on either side of the barrier.  The partially-wet state has a macroscopic liquid layer on one side of the barrier; the other side is coated by a thin film of finite width $\zeta$, see also Sections \ref{sec:background} and \ref{subsec:wetting_transition}. {(Here, the term ``thin'' indicates that the layer width is comparable to the interfacial width, i.e. microscopic;  it is much smaller than the system size.)}  The unpinned {(dry)} state is defined by the liquid bulk detaching from the barrier entirely; in contrast to systems with noise \cite{Turci2021}, this {drying transition} is only observed when the barrier is removed completely, i.e. on the $\epsilon=0$ line\footnote{We note that this line is labelled ``unpinned'' to illustrate this point, although the physics at $\epsilon=0$ is that of standard MIPS.}. This may be expected, since any repulsive barrier causes accumulation above the vapour density; it is then always favourable to have this density accumulation inside a liquid phase or near a liquid-vapour interface (such that it overlaps with the exponential tails of this interface). Any barrier therefore breaks translational symmetry, which is sufficient to localise the liquid slab in our noise-free dynamics. The homogeneous state occurs for $\text{Pe}< \text{Pe}^* = 4.33$ where the system does not phase separate and wetting phenomenology is not expected. 

Full wetting occurs for barriers larger than the critical barrier height $\epsilon^*(\text{Pe})$, above which the wall is effectively ``hard'' \cite{Neta2021, Turci2021, Das2020}. Additionally, Figure \ref{fig:phase_behaviour}(a) shows this critical barrier height $\epsilon^*(\text{Pe})$ is \textit{increasing} with the activity $\text{Pe}$. This is also unsurprising; at higher activity, particles exert a stronger driving force and can more easily overcome the barrier. The barrier thus becomes effectively softer and more permeable, favouring partial wetting (recall Section \ref{sec:choice_permeable_barrier}).

We check for finite size effects by repeating this procedure with a larger $\ell_s$ at isolated points on the surface phase diagram. We do not observe any structural changes to the transition lines or significant movements of their positions, and so we conclude that the surface phase-diagram is well-converged to its large-system limit. The boundary between full and partial wetting in Figure \ref{fig:phase_behaviour}(a) is obtained numerically by testing the linear stability of the fully-wet state (see Section \ref{subsec:numerical_implementation}).  We do not observe any metastability at this transition (see Section \ref{subsec:wetting_transition} below), so this procedure allows accurate determination of the phase boundary based on short simulations. 

The steady-state results for $f(x,\theta,t)$ are consistent with expectations for active systems \cite{Reichhardt2017}, in that the orientations of the active particles are isotropic in bulk phases but develop anisotropic structure at the liquid–vapour interface and at the barrier.   One observes significant probability currents in steady-state; these are divergence-free (consistent with $\partial f/\partial t=0$) and arise due to the particles' self-propulsion.

\subsection{The wetting transition is critical}\label{subsec:wetting_transition}

We now examine in detail the wetting transition, which separates fully and partially-wet states. 
We find that this transition bears all the characteristics of an equilibrium critical wetting transition (recall Sec.~\ref{sec:background}).  The control parameter for active wetting is the barrier height $\epsilon$. 
Figure \ref{fig:wetting_transition} shows results for $\text{Pe}=10$. 
We verified that the required qualitative features are robust across a range of $\text{Pe} > \text{Pe}^*$, and for other values of $\bar{\rho}$ inside the spinodal. The features are:

\textit{(a) Continuous divergence of the liquid film width.}  Recall from \eqref{eq:log_div} that for critical wetting, the dry wall in a partially-wet state carries a thin liquid film whose width $\zeta$ diverges continuously at the wetting transition.  Evidence for such a divergence is shown in Figure \ref{fig:wetting_transition}(a), which may be contrasted with a first-order wetting scenario, where $\zeta$ would have a finite limit (and finite gradient) as the transition is approached.  The nature of the divergence is logarithmic in equilibrium \cite{Cahn1977, DeGennes1985} but we are not aware of any theoretical predictions for active systems.

\textit{(b) Square-root scaling of the asymmetry.} 
The fully-wet state is symmetric under reflection about $x=0$ so ${\cal A}=0$, while the partially-wet state has ${\cal A}>0$.  Recall from \eqref{eq:Aeq-sqrt} that for critical equilibrium wetting, the asymmetry is predicted to vanish as $(\epsilon^*-\epsilon)^{1/2}$ on approaching the transition.  This behaviour is illustrated in Figure \ref{fig:wetting_transition}(b); the exponent $1/2$ appears robust even out of equilibrium, as may be expected generically for pitchfork bifurcations.

\textit{(c) Vanishing hysteresis loop.} Figure \ref{fig:wetting_transition}(b) also contrasts the behaviour on increasing and reducing $\epsilon$ across the transition.  There is no evidence for metastable states (neither partially- nor fully-wet).  This is the expected behaviour for critical wetting. 
At a first-order transition, we would expect our noiseless hydrodynamic equations to instead support a finite hysteresis loop, even if $\epsilon$ is varied quasi-statically.
  
\textit{(d) Linear instability of the fully-wet state.} Another characteristic feature of critical wetting, which provides further confirmation of the absence of metastability, is that the fully-wet state becomes linearly unstable to density perturbations exactly at $\epsilon^*$. This means that
\begin{equation}
\dv{t} \ln(\mathcal{A}(t)) = \lambda (\epsilon^* - \epsilon) + \mathcal{O}((\epsilon^* - \epsilon)^2).
\label{equ:instab}
\end{equation}
This behaviour is shown numerically in Fig.~\ref{fig:wetting_transition}(c): we apply a small perturbation to the fully-wet state and measure the asymmetry as a function of time. As expected, we see linear instability of the fully-wet state below the critical barrier height $\epsilon^*$.

The transition between fully- and partially-wet states therefore bears all the hallmark characteristics of a critical wetting transition, and so we conclude that this active wetting transition is critical{, at least in a two-dimensional mean-field-type model -- see also \cite{Turci2021}}.
In the Cahn theory of equilibrium wetting, critical wetting occurs when the effect of the barrier in preventing attractive interactions between fluid particles is significant (i.e. $c_2$ is large, recall Section \ref{sec:background} and Appendix \ref{app:wetting}). In the active wetting scenario, these attractions are indeed strongly suppressed near the barrier; the large polarisation $\vb{p}$ near the barrier means that active particles near the barrier are locally aligned. Since effective attractive interactions in MIPS arise due to collisions between oppositely aligned particles~\cite{Cates2015}, the barrier inhibits the attractive fluid-fluid interactions dramatically -- in the large Pe limit, we expect this inhibitive effect to become absolute. This offers a physical explanation for the observation that the wetting transition {at two-phase coexistence} is critical in the active system.

\subsection{Ratchet currents in partial wetting}
Although the steady-states are qualitatively similar to their passive counterparts, nonequilibrium consequences of activity are also visible. In particular, we show in \cite{Grodzinski2026a} that any asymmetrically wet barrier (i.e. $\rho(x) \neq \rho(-x)$ for $x \in [-w_b, w_b]$) must have a steady-state (uniform) current $J^{(\rho)}$ flowing through it. This ratchet current flows through the barrier from the ``dry'' (low-density) to the ``wet'' (high-density) side (i.e. in the opposite direction to diffusion). 

\begin{figure}
    \centering
    \includegraphics[width=1.0\linewidth]{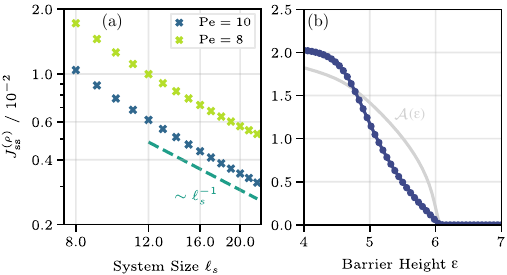}
    \caption{Steady-state ratchet current in the partial-wetting scenario. (a) System size dependence, asymptotically going to $1/\ell_s$. (b) Continuous vanishing of the steady-state current as the wetting transition is approached (same parameters as Figure \ref{fig:wetting_transition}(b) -- asymmetry also shown).}
    \label{fig:current}
\end{figure}

Although ratchet currents are expected for any system of (non-momentum-conserving) active particles in a potential that violates reflection symmetry \cite{Reichhardt2017, Metzger2025}, the current here is a result of spontaneously broken mirror symmetry about a symmetric barrier. This spontaneous ratchet current therefore lies beyond the usual \textit{ratchet principle} \cite{Reichhardt2017}.
Since this current arises from a localised driving force on the barrier, while friction acts against all particles, we expect the current to scale as $J^{(\rho)} \sim \ell_s^{-1}$, and also to vanish continuously as the (critical) wetting transition is approached. This behaviour is confirmed in Figure \ref{fig:current}.

This ratchet current has two consequences. First, a linear density gradient is present throughout the bulk liquid\footnote{A linear gradient should also develop over the vapour bulk for the same reason; however, this is less evident in the numerical steady-states due to the longer tails of the liquid-vapour interfaces into the vapour bulk.}, leading to an $\mathcal{O}(1)$ deviation of the density from its binodal value $\rho_l$, that persists far from any interfaces (see Figure \ref{fig:phase_behaviour}(c), magnified section). We verified that the density drop across the liquid is asymptotically constant as $\ell_s \to \infty$ (as expected, see discussion in \cite{Grodzinski2026a}), and so this effect is apparent even in the limit of a large system\footnote{{We emphasise that this density deviation is distinct from the corrections to the density of a thin adsorbed film that arise in equilibrium wetting due to interactions between the surface-liquid and liquid-vapour interfaces. The active density shift is a bulk effect: the linear density gradient extends throughout the entire bulk layers, as a non-local consequence of the ratchet current.}}. 

Second, the presence of a current through the barrier leads to a novel dynamical pathway for the full-to-partial wetting transition. Specifically, this transition involves the ``pinching off'' of one liquid layer, leading to a detached slab of liquid on one side of the barrier, which evaporates gradually until a partially wet state is reached. Such a transition would be forbidden in a passive fluid due to the creation of new interfaces; the equilibrium pathway (which proceeds due to interactions between the liquid-vapour interface and the wet barrier) is much slower. A more detailed account of this novel dynamical pathway is presented in \cite{Grodzinski2026a}, which also provides a simplified mechanistic derivation in terms of the global effects of an interfacially localized, active pump.

\section{Discussion}\label{sec:discussion}
\subsection{Summary of results}\label{subsec:summary}
We have studied active wetting by considering the hydrodynamic behaviour of an active lattice gas in the presence of a smooth permeable barrier, under periodic boundary conditions. This geometry is analogous to a slit geometry widely used to study equilibrium wetting \cite{Evans2019, Leeuwen1989, Nijmeijer1990, Sikkenk1987}. We take as control parameters the barrier height $\epsilon$ and self-propulsion $\text{Pe}$.

The surface phase diagram reveals wetting states whose density profiles are analogous to the equilibrium wetting states known as dry, partially-wet, and fully-wet. As our model is hydrodynamic (i.e. large-scale) and is studied under conditions of liquid-vapour coexistence, we conclude that these  are true macroscopic wetting states, instead of microscopic accumulations of active matter or active adsorption. We observe a drying transition at $\epsilon=0$ and a wetting transition at finite $\epsilon = \epsilon^*(\text{Pe})$. The wetting transition displays all the salient characteristics of equilibrium critical wetting \cite{Cahn1977, DeGennes1985, Bonn2009}, including a diverging film width on the dry side, vanishing hysteresis, and a pitchfork bifurcation in the asymmetry order parameter $\mathcal{A}$.

Our hydrodynamic framework also facilitates an investigation of new physical phenomena in active wetting, which arise as a consequence of activity. We demonstrate two pronounced nonequilibrium effects. First, the permeable barrier acts as a ratchet, pumping mass from the ``dry'' to the ``wet'' side of the wall. This leads to a circulating current in the partially-wet steady-state (recall Sec.~\ref{subsec:active_wetting_states}) and departure of the bulk densities from their binodal values. Second, we show that this ratchet current gives rise to a novel dynamical pathway between fully- and partially- wet states, which is inaccessible for passive fluids, as discussed in greater detail in \cite{Grodzinski2026a}. Overall, these findings establish a precise correspondence between active and equilibrium wetting, while allowing us to identify subtle yet significant nonequilibrium phenomena which arise solely due to activity. 

Our results complement previous findings from numerical investigations of active wetting. We use the same geometry as in \cite{Turci2021} and \cite{Das2020}, i.e. a slit geometry with periodic boundaries and a permeable wall; however, our deterministic equations allow us to analyse scaling in the wetting transition without finite-size effects (albeit in a model of mean-field-type).  We find that wetting in two dimensions is critical, consistent with~\cite{Turci2021, Neta2021}. Finally, we emphasise that the connections we find between active and equilibrium wetting arise when the active system is in a state of two-phase coexistence. This is distinct from previous work studying accumulation/adsorption of particles on boundaries, in single-phase active fluids~\cite{Das2018, Elgeti2013}.

\subsection{Outlook}\label{subsec:outlook}

There are several promising directions for future work. The introduction of translational symmetry breaking in the direction(s) parallel to the wall would allow investigation of other partially-wet geometries (i.e. ``droplet'' and ``bridge'') within the same model. The presence of droplets on the barrier in the partial wetting state could result in a circulating flux pattern on the macroscopic scale, even with closed boundary conditions. Additionally, extension of this work to two dimensions would allow a comparison with the results of \cite{Zhao2026}, in which the authors identify local steady-state currents around the contact line as a robust feature of active partial wetting which modifies contact angles and droplet shapes. Our deterministic hydrodynamic model would allow the theoretical predictions contained in this work to be explored in greater detail, without finite-size effects, but subject to mean-field scaling. 

Another extension would be to introduce dynamical noise to the system, which can be done in our lattice-based approach in a controlled manner via a leading-order correction to the hydrodynamic equations \cite{Agranov2021, Martin2025}, or by using a noisy coarse-grained field theory such as Active Model B+ \cite{Tjhung2018}. The inclusion of dynamical noise and fluctuations can have dramatic consequences in equilibrium \cite{Evans2019} and nonequilibrium \cite{Turci2021} wetting phenomenology, including changing the stability of steady-states and the order/scaling of transitions. In active systems, noise leads to a wide variety of ``exotic'' phase-separated states \cite{Cates2025} (these are not seen here due to the hydrodynamic limit, see also \cite{Mason2023a, KourbaneHoussene2018}). Investigating the consequences of noise in active wetting, and the associated bubbly phase separated states, could therefore help confirm the generality and robustness of the findings presented here. 

Finally, it is important to consider how the effects presented here might be observed or harnessed experimentally. A promising experimental direction could be to recreate active wetting in a slit geometry using light-activated colloidal particles \cite{Vutukuri2020, Rey2023, Stenhammar2016}, in which activity can be spatially confined to the central region of a slit without the need for confinement in the perpendicular direction.
This approach (which allows for quasi-one-dimensional periodic boundary conditions, for example by global confinement in a toroidal region, into which a force-field of barrier type could also be introduced optically~\cite{Stenhammar2016}) may enable direct measurement of asymmetries, currents, and wetting states, and could provide direct experimental evidence of the nonequilibrium effects described here and in \cite{Grodzinski2026a}.

\begin{acknowledgments}
We thank Robert Evans, Francesco Turci, Maria Bruna, Nigel Wilding, and Uwe Thiele for insightful discussions and valuable feedback throughout this work. This work was supported by Engineering and
Physical Sciences Research Council grants number EP/W524633/1 (Project Reference 2927750) and EP/Z534766/1.
\end{acknowledgments}

\section*{Data Availability}
The code and data that support our results are openly available at~\cite{Grodzinski2026b}.

\appendix

\medskip
\section{EQUILIBRIUM THEORY OF CRITICAL WETTING}
\label{app:wetting}

\subsection{Cahn theory of wetting}
\begin{figure}[t]
  \centering
  \includegraphics[width=8cm]{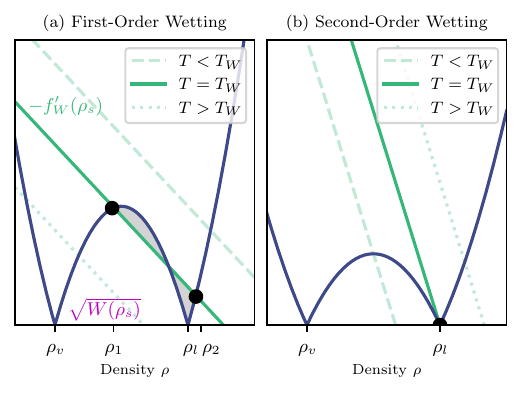}
  \caption{Illustration of First-order and Critical (second-order) wetting in the Cahn theory. 
  (a) In a first-order wetting transition, $c_2$ is small, and multiple solutions for $\rho_s$ exist; the discontinuous transition between the solutions is determined by a Maxwell condition.
  (b) In a critical transition, only one solution exists for $\rho_s$, and the wetting transition occurs when this passes continuously through $\rho_l$.}
  \label{fig:Cahn_Constriction}
\end{figure}

We briefly review some of the results of the Cahn theory of wetting, as discussed in \cite{Cahn1977, DeGennes1985}. These are useful points of comparison with the active wetting model, as summarised in the main text. 
Minimisation of the free energy \eqref{F_eq} results in a condition on the fluid density at the wall $\rho_s \equiv \rho(0)$:
\begin{equation}
  -\dv{f_w}{\rho_s}= \sqrt{2 \kappa W(\rho_s)}.
\end{equation}
Here, without loss of generality, we have assumed that the bulk free-energy density $W(\rho)$ has minima $W(\rho_l) = W(\rho_v) =0$. This may give one or multiple solutions for the surface density $\rho_s$; stable solutions are shown by the black points in Figure \ref{fig:Cahn_Constriction}.
If the solution $\rho_s \in [\rho_v, \rho_l)$, the density profiles approach $\rho_v$ as $x \rightarrow \infty$ (in the bulk) to form a partially-wet/dry wall, or approach $\rho_l$ to form a fully-wet wall. If $\rho_s \geq \rho_l$, decay to the liquid density in the bulk is always energetically favourable, and the wall is always fully wet.

In a first-order wetting transition, two stable solutions exist for $\rho_s$, with $\rho_1 \in (\rho_v, \rho_l)$ and $\rho_2 > \rho_l$. The first solution $\rho_1$ corresponds to partial wetting, and the second solution $\rho_2$ to full wetting (see Figure \ref{fig:Cahn_Constriction}(a)). The transition point between these is determined by a Maxwell construction \cite{DeGennes1985}. The surface density changes discontinuously, and metastable solutions for $\rho_s$ exist. 

Critical (second-order) wetting occurs when there is only one solution for $\rho_s$ near the wetting transition, i.e. if $c_2 > \lim_{\rho \rightarrow \rho_l} \frac{\partial}{\partial \rho}[\sqrt{2 \kappa W(\rho)}]$. In this case, $\rho_s$ passes continuously through $\rho_l$ and no metastable states exist.

Just below a critical wetting transition, we have $\rho_s = \rho_l - \delta$. By minimising the free energy of the density profile away from the wall, we find $\delta \propto e^{-\zeta/\xi}$, where $\zeta$ is the thickness of the wetting film and $\xi$ is the liquid-vapour interfacial width~\cite{DeGennes1985}. Taking temperature as a control parameter, the free energy of this film is then given by (for a detailed discussion, see \cite{DeGennes1985}):
\begin{equation}\label{eq:curly_F}
    F_\text{film}(\zeta) = F_0 - \alpha(T) e^{-\zeta/\xi} + \beta(T) e^{-2\zeta/\xi},
\end{equation}
where $\alpha \propto (T_w - T) + \mathcal{O}((T_w - T)^2)$, and $\beta = \text{constant} + \mathcal{O}(T_w - T)$. The effect of temperature $T$ is to control the shape of $W(\rho)$; at a higher temperature, the coexistence densities become closer and the energetic barrier between these densities becomes smaller. The optimal width of the liquid film $\zeta^*$, which minimises the free energy \eqref{eq:curly_F}, is thus given by:
\begin{equation}\label{eq:log_div-app}
  \zeta^* = -\xi \log(T_w - T) + \text{const.}
\end{equation}
The width of a liquid film near a critical-wetting transition therefore diverges logarithmically.
In contrast, the film width at a first-order wetting transition will jump discontinuously between microscopic ($\sim \xi$) and macroscopic values.
A similar analysis demonstrates that the divergence of the liquid film width is logarithmic approaching a critical transition for any other control parameter (such as the wall attractiveness $c_1$, as in the main text).

In a slit geometry, it is favourable for two initially equal liquid films of width $\zeta_0$ to remain equal in size (so the system will remain fully wet) if and only if:
\begin{equation}\label{eq:concavity}
    \eval{\dv[2]{F_\text{film}}{\zeta}}_{\zeta_0} \leq 0,
\end{equation}
which, for $\zeta_0 \rightarrow \infty$, occurs precisely when $T \geq T_w$. Below this critical temperature, it is favourable for mass to redistribute from one film to another (via a translational instability) until one film is microscopic ($\sim \xi$) and the other is macroscopic; this is the partial wetting scenario.
A fully-wet state is therefore unstable below this critical transition temperature in a slit geometry, and a partially-wet state is unstable above it.
Thus, there are no metastable states in the critical wetting scenario, as mentioned in the main text. Once again, this analysis is general for any other choice of order parameter.

\vspace{1em}
\subsubsection{Asymmetry order parameter}\label{Appendix_Asymmetry}
The Cahn theory concerns the behaviour of a fluid near a single, attractive boundary. We now derive the behaviour of a fixed total mass of fluid in a slit geometry (i.e. in the presence of two barriers) from the Cahn theory, for comparison with the active case.

Consider wetting in a slit geometry of total width $L_\text{tot} > 2L$, with two liquid films of thickness $L-\delta, L+\delta$ on either wall (and without loss of generality $\delta \in [0, L]$).
We assume that $L \gg \xi$, such that the liquid-vapour interfaces are microscopic.
The free energy due to the wetting layers becomes:
\begin{multline}
   F_\text{film}(\delta) = - 2 \alpha(T) e^{-L/\xi} \cosh(\delta / \xi) \\ + 2 \beta(T) e^{-2L/\xi} \cosh(2\delta / \xi).
\end{multline}
To find the optimal $\delta$, we set $\pdv{F_\text{film}}{\delta} = 0$, yielding the condition:
\begin{equation}
    \cosh(\delta / \xi) = \frac{\alpha}{4 \beta} e^{L/\xi}.
\end{equation}
As the liquid film thickness varies continuously approaching a critical wetting transition, we expect $\delta$ to be small just below criticality. So expanding $\cosh(\delta/\xi) \simeq 1 + \frac{1}{2}(\frac{\delta}{\xi})^2$:
\begin{equation}
   \delta + \mathcal{O}(\delta^2) = \xi \sqrt{\frac{\alpha}{2 \beta} e^{L/\xi} - 2} .
\end{equation}
The asymmetry parameter can be expressed as $\mathcal{A} = \frac{L^+ - L^-}{L^+ + L^-} = \frac{\delta}{L} \in [0, 1]$ (this is equivalent to \eqref{eq:Aeq-sqrt}). Define $\bar{\alpha} = \alpha - 4 \beta e^{-L/\xi}$. We may assume that this is proportional to a modified critical temperature near criticality, i.e. $\bar{\alpha} = \mu (\bar{T}_w- T) + \mathcal{O}(|\bar{T}_w - T|^2)$. 
The modified critical temperature accounts for a finite system size correction: $\bar{T}_w = T_w + \mathcal{O}(e^{-L/\xi})$.
With these definitions, we have the following behaviour for $\mathcal{A}$ near a critical wetting transition, in a finite system:
\begin{equation}
\label{eq:pfork}
    \mathcal{A} =  
\begin{cases}
0 & \quad T \geq \bar{T}_w, \\
\left[ \frac{\xi}{L}  \sqrt{\frac{\mu}{2 \beta}} e^{\frac{L}{2\xi}} \right] \,  (\bar{T}_w - T)^{1/2}  & \quad T < \bar{T}_w. 
\end{cases}
\end{equation}
{One sees that this asymmetry exhibits a pitchfork bifurcation near a critical wetting transition, with characteristic exponent $1/2$. 
In the large-system limit $L\to\infty$ then ${\cal A}\to \mathcal{O}(1)$ for all $T<T_w$; this effect is signalled in Eq.~\eqref{eq:pfork} by the divergence of the prefactor in this limit.  In practice, this means that observation of power-law scaling in Eq.~\eqref{eq:pfork} requires relatively small system sizes ($L/\xi \sim 1$), and $T$ close to $\bar T_w$.}

\subsection{Categorisation of Surface Transitions}\label{app:wetting_categories}
{To contextualise the wetting and surface accumulation behaviours observed in our model, we provide a brief review of the corresponding wetting transitions and surface accumulation phenomena in equilibrium. In particular, we distinguish between critical wetting, complete wetting, adsorption, layering, and prewetting. While the active wetting literature has not always strictly maintained these distinctions, this framework is necessary for interpreting our results and comparing them with their equilibrium counterparts. For more comprehensive reviews, see, e.g., \cite{Sullivan1986, Dietrich1988, Binder1988, Nakanishi1982, DeGennes1985, Schick1990}.}

{In equilibrium (for example \cite{Binder1988}), wetting may be studied using lattice models such as a semi-infinite Ising system. This model is parameterised by a bulk field $H$ (setting the bulk chemical potential), a bulk coupling constant $J$, and the temperature $T$. The surface introduces a planar edge layer governed by a surface field $H_1$ (controlling the short-range wall-fluid interaction), and a modified surface coupling $J_s$, where the ratio $J_s/J$ quantifies how the wall alters local fluid-fluid interactions.
}

{The study of wetting focuses on the behaviour of the interface between the phase adsorbed at the wall and the bulk phase. Depending on the thermodynamic path taken, several distinct surface phenomena emerge (summarised in Table \ref{tab:wetting_phenomena}) \cite{Lipowsky1982, Lipowsky1984}:}

\begin{itemize}
    \item \textbf{Adsorption:} A generic, finite accumulation of fluid at any attracting surface, occurring without a phase transition.
    
    \item \textbf{Critical and First-Order Wetting Transition:} Surface phase transitions that occur exactly at bulk coexistence ($H = 0$). Tuning the surface field $H_1$ toward a critical value can cause the initially microscopic adsorbed film thickness to diverge to a macroscopic value, either continuously (critical wetting) or discontinuously (first-order wetting).
    
    \item \textbf{Complete Wetting Transition:} This describes the behaviour \textit{after} the surface parameter is already wet, but before coexistence. The (unbounded) film grows continuously as the bulk field $H$ approaches coexistence ($H \to 0^+$).
    
    \item \textbf{Layering Transitions:} At temperatures below the roughening transition ($T < T_R$), the discrete lattice geometry (or packing geometry) becomes relevant. Wetting then proceeds via a series of discrete layer-by-layer filling transitions rather than continuous growth.
    
    \item \textbf{Prewetting Transition:} When the wetting transition at coexistence is first-order, the discontinuity in the film thickness can extend off-coexistence into the bulk region ($H \neq 0$). This extension is known as a prewetting transition, where the film thickness jumps from thin to thick before eventually diverging at bulk coexistence. The prewetting line ultimately terminates at a surface critical point.
\end{itemize}

\begin{table*}
\centering
\begin{tabular}{l c c l}
\hline\hline
\textbf{Phenomenon} & \textbf{Fixed} & \textbf{Tuned} & \textbf{Features} \\
\hline
Adsorption & & & Finite surface excess\\[4pt]
Critical wetting & $H=0$ & $H_1 \to H_{1c}$ \quad & $m_s \sim -\ln|H_1 - H_{1c}|$ \\[4pt]
Complete wetting & $H_1 < H_{1c}$ & $H \to 0^+$ & $m_s \sim -\ln(H/J)$ \\[4pt]
Layering & $T < T_R, H=0$ \quad & $H_1$ & Discrete filling \\[4pt]
\hline\hline
\end{tabular}
\caption{{Summary of distinct surface phenomena in the Ising wetting problem (following Binder \& Landau, \cite{Binder1988}). Critical wetting is a two-field surface phase transition at coexistence; complete wetting is a one-field divergence within the wet phase; adsorption is a generic accumulation. The surface excess $m_s=\sum_{\text{layers } n} (m_\infty - m_n)$ measures the total deviation of the magnetisation from its bulk value.}}
\label{tab:wetting_phenomena}
\end{table*}

\begin{figure}[t]
    \centering
    \includegraphics[width=1.0\linewidth]{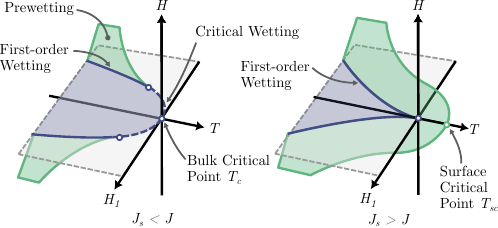}
    \caption{Surface phase diagram for the wetting and prewetting transitions for $J_s<J$ (left), $J_s>J$ (right); after \cite{Nakanishi1982, Binder1988}. First-order transitions are shown with bold lines; critical transitions are shown with dashed lines; multicritical points are shown with open circles.  The blue region is the fully-wet area of the phase diagram, while the green region indicates prewetting (which only appears attached to a first-order wetting transition).}
    \label{fig:eqm_wetting_PD}
\end{figure}

{The topology of the surface phase diagram depends on the ratio $J_s/J$. Importantly, critical wetting can only occur for $J_s/J \leq 1$. For $J_s/J>1$, a surface critical point appears at temperature $T_{sc}$ (distinct from the bulk critical temperature $T_c$), associated with prewetting. (See Figure \ref{fig:eqm_wetting_PD} for an illustration of the surface phase diagram.)}

{The same categorisations also refer meaningfully to wetting in off-lattice models, and in models with long-range interactions. Although exact results are typically unavailable for models containing long-range fluid-fluid or fluid-wall interactions (e.g. Van der Waals forces), as well as fluctuations, DFT simulations still allow these cases to be studied with precision, using the same classification of surface phase transitions \cite{Evans2019}. }

{The parameters in our model have a clear qualitative analogy to this equilibrium wetting model. The barrier height  $\epsilon$ plays the role of $H_1$, while $\text{Pe}$ dictates the effective fluid-fluid interactions, analogous to $J$. By fixing the global density $\bar{\rho}$ between the binodal and spinodal lines, we work strictly at bulk MIPS coexistence, analogous to $H = 0$. Furthermore, the limits taken to derive our hydrodynamic description naturally preclude discrete layering transitions. Finally, as noted in Section \ref{subsec:wetting_transition}, the orientational alignment of active particles near a wall tends to inhibit their effective attractive interactions. Consequently, our active model is analogous to the equilibrium scenario where $J_s < J$, illustrated in Figure \ref{fig:eqm_wetting_PD} (left). As the only microscopic inter-particle interactions are short-ranged steric repulsion, and the wall potential is truncated, there are no long-range forces in our model.}

\subsection{Slabs and Droplets in Partial Wetting}
\label{app:slabs_droplets}
{
In the partial-wetting regime of a slit geometry in two dimensions (i.e. $\cos(\theta_w) \in (-1, 1)$), the liquid bulk can adopt three qualitatively distinct ``morphologies'', or shapes \cite{Binder2003, Swain2000, Turci2024}: 
\begin{itemize}
    \item \textbf{Slab}: A laterally uniform liquid bulk coating one wall (with a microscopically thin film on the other ``dry'' wall), as in Figure \ref{fig:wetting_schematic}(b.2).
    \item \textbf{Droplet(s)}: One or more circular segments of liquid bulk on the wall (as in Figure \ref{fig:wetting_schematic}(a)), with a thin film on the remaining area of wall.
    \item \textbf{Bridge}: A bridge of bulk liquid connecting the two walls of the slit, whose interfaces are arcs of circles.
\end{itemize}

For thermodynamic ground states of passive systems, liquid-vapour interfaces are circular arcs and contact angles obey Young's equation.  The observed morphology depends on four parameters; the slit aspect ratio $L_y/L_x$, two distinct interfacial tensions (where we may set $\gamma_\text{LV} \equiv 1$) that set the contact angle $\theta$, and the liquid volume fraction $\phi$.

The existence of some morphologies depends on geometric constraints. For example a liquid droplet is only possible if it physically fits inside the slit.  Among the existing morphologies, the ground state is the one that minimises the free energy, which (at bulk coexistence) corresponds to minimising the lengths of interfaces multiplied by their relevant surface tensions. (Any metastable state which is not the ground state will eventually transition to the ground state via a rare fluctuation.) Using the above considerations, a straightforward numerical calculation reveals that droplet states are preferred for small liquid volume fractions, while slab states are preferred if the aspect ratio is small, and bridge states are preferred if the slit aspect ratio is large.

Extending these arguments to active matter systems is difficult in general. For example, the presence of steady-state currents means that droplet interface shapes and contact angles cannot be easily predicted \cite{Zhao2026}.  This means that even the geometrical constraints are not straightforward to enforce.  Moreover, the absence of a global free energy means that metastable and globally-stable ground states cannot be distinguished in hydrodynamic theories, although stability to weak perturbations remains clear. Slab \cite{Turci2021}, droplet \cite{Das2020}, and bridge \cite{Zhao2026} states have all been observed in active wetting systems in a slit geometry.

The one-dimensional calculations of active wetting in this paper describe only the slab states.  We certainly expect that these should be the dominant states for small aspect ratios, in which droplets and bridges both tend to be suppressed by geometrical constraints.  More generally, we at least expect these slab states to be metastable.
We have verified this numerically for several points in the phase diagram of Figure \ref{fig:phase_behaviour}(a): we initialised partially-wet slab states in a two-dimensional realisation of our system and added small, spatially uncorrelated perturbation to $f$.  These perturbations decay under the hydrodynamic time evolution, confirming the stability of the slab. Future work will investigate the behaviour of droplet and bridge states.
}

\section{NUMERICAL IMPLEMENTATION}\label{app:timestepping}
\subsection{Numerical Timestepping Routine}
To simulate the dynamics of the model, as well as generate the steady-states, we use an explicit time-stepping routine with one spatial and one angular dimension. We first rewrite the equations of motion in the form:
\begin{equation}
  \partial_t f = - \partial_x [M^x U^x] - \partial_\theta [M^\theta U^\theta] ,
\end{equation}
where $M^x, M^\theta$ are the scalar mobilities and $U^x, U^\theta$ are the velocity fields. The mobilities are given by $M^x = f d_s(\rho)$, $M^\theta = f$, and the velocity fields by:
\begin{multline}\label{eq:U^x}
    U^x = - D_E \left[ \partial_x (\log{f}) + \frac{\mathcal{D}(\rho)}{d_s(\rho)} \partial_x \rho + \frac{ (1-\rho)}{d_s(\rho)} \partial_x V\right] \\ + v_0 \left[ \cos(\theta) + \frac{p_x s(\rho)}{d_s(\rho)} \right]
\end{multline}
and 
\begin{equation}\label{eq:U^theta}
    U^\theta = - D_O \partial_\theta [\log{f}].
\end{equation}
where $p_x \equiv \int_0^{2\pi} f(x,\theta) \cos(\theta) \dd{\theta}$ is the x-component of the polarisation. We then follow the same first-order update scheme as Section 4 of \cite{Mason2023a} and Appendix E of \cite{Mason2025}, but modified with the slightly different angular and spatial dynamics \eqref{eq:U^x}-\eqref{eq:U^theta}. We discretise the system into an $N_x \times N_\theta$ grid, and use an upwind condition to calculate the fluxes, updating the discretised field $f$ accordingly. A Courant–Friedrichs–Lewy (CFL) condition is used for an adaptive timestep\footnote{Although the CFL condition is necessary to prevent instabilities/inaccuracy arising from advective behaviour, it does not address issues related to the density leaving its valid range of $\rho \in [0,1]$. For this reason, $a, b > 1$ and finite $\tau_\text{max}$ are necessary to ensure stability.} (as in \cite{Mason2025, Mason2023a}):
\begin{equation}
    \Delta t_\text{max} = \min \left( \frac{\Delta x}{a \,  \max \abs{U^x}}, \frac{\Delta \theta}{b \,  \max \abs{U^\theta}}, \tau_\text{max} \right).
\end{equation}
We find that parameters $a=b=6, \tau_\text{max} = 10^{-4}$ are sufficient to ensure numerical stability and accuracy, while remaining practical to compute. We typically ran simulations with $N_x = 200$ and $N_\theta = 30$, although we verified that results are accurate when compared with finer grids. We always use an even $N_\theta$ to ensure the simulation is parity-symmetric.

Various checks were performed to verify that this integration scheme functions as expected; for example, the mean density is conserved up to a computational error that is typically of order $10^{-10}$, the averages of higher angular moments vanish similarly in steady-state, and the stability boundary matches the theoretical prediction of \cite{Mason2023a} (see Figure \ref{fig:binodal_spinodal}). Symmetry-breaking events were confirmed to be random in direction, as expected -- for example, in a trial of 70 partially-wet states, the liquid layer formed on the left of the barrier in 37 simulations. The dynamics of full wetting was qualitatively compared to the particle-level simulations of \cite{PerezBastias2025} and \cite{Turci2021}, confirming that the hydrodynamic equations reproduce the appropriate physical behaviour.

\subsection{Generation of binodal and spinodal}\label{sec:binodal_spinodal_convergence}

For the bulk fluid (without boundaries), the presence of two-phase coexistence is characterised by the spinodal (which is the boundary of linear stability of the homogeneous fluid) and the binodal (which determines the densities of the coexisting liquid and vapour).
To compute the spinodal, we follow the method outlined in Section 3 of \cite{Mason2023a}. As the spinodal is defined in the absence of a potential, our modified equations have no effect on the linear stability analysis. For the theoretical spinodal line shown in Figure \ref{fig:binodal_spinodal}, we truncate the (infinite) instability matrix of a homogeneous state to a $10 \times 10$ matrix, which is sufficient for good convergence as discussed in \cite{Mason2023a}.

We confirm this theoretical prediction by finding the spinodal numerically.  For this, we initialise the system with a homogeneous density field, and then add a small perturbation corresponding to the eigenfunction with the largest eigenvalue (if this eigenvalue is positive, we expect instability). We then classify the phases according to the mean square deviation of the density field from the homogeneous state, using the same classification scheme as in \cite{Mason2023a}. Although the spinodal is formally defined in the limit of large system size, our simulations are run in a system size of $L_x = 20$ with a grid of $200 \times 30$ points.
We find good agreement between theory and numerics, as shown in Figure \ref{fig:binodal_spinodal}. No significant deviations are visible if a larger system size or smaller lattice spacing are used. The theoretical spinodal also provides a check for the validity of our numerical routine and computational parameters, as we can see that the physics of stability are preserved appropriately.

A theoretical expression for the binodal is not available, so we must rely on numerical methods to generate it.
We use the following routine:
\begin{enumerate}
    \item Initialise a simulation from a homogeneous state plus a maximal instability perturbation as above. 
    The parameters used are: $L = 20, N_x = 400, N_\theta = 30, \rho = \rho^* = 0.6585$, and $v_0$ is a specified parameter. If $v_0 > 15$, we instead use $N_x = 800$ so the (sharper) interfaces remain well-resolved.
    \item Evolve this system to time $t_0 = 200$, after which a single liquid slab has formed.
    \item Initialise a new simulation with the same parameters apart from $L$, but using reflecting boundary conditions (i.e. $\eval{\dv{f(x, \theta)}{x}}_{x=0, L} = 0$), in a system of size $L = 40$. 
    Take one half of the previous simulation as a new initial condition, such that the profile contains vapour on one side and liquid on the other, with a single interface.
    \item Evolve to a time $t_1 = 200$. Identify the liquid and vapour densities as the maximum and minimum densities in this final state.
    \item Repeat for 8 uniformly spaced values of $v_0 \in (v_{0, \text{crit}}, 6.0]$ and another 10 values in the range $v_0 \in (6.0, 30.0]$.
\end{enumerate}

The numerical binodal is then generated by interpolating between the discrete data points found via steps 1-5 above. The effective size over which the interface is resolved is $L=40$, and so we expect convergence to the bulk values given a typical interfacial scale $l_D = 1.0$, and approximately exponential tails of the liquid-vapour interface.
Simulations were run for isolated points on the binodal with a larger system size and finer grid spacing; no significant departure was seen from the interpolated binodal, and so we assume that the above parameters are sufficient for good convergence for the range of Pe shown.
For example, the binodals for $v_0 = 10.0$ were estimated at $(\rho_v, \rho_l) = (0.1086, 0.9855)$ and $(\rho_v, \rho_l) = (0.1115, 0.9872)$ for systems of size $L=40, L=50$ respectively, a difference of about 0.2\%.
Similar checks were performed for several values of $v_0$, and all quoted values did not change by more than 1\% when the system size was increased to $L=50$. Therefore, we are confident that the binodal is sufficiently converged for the purposes of the figure.

\subsection{Classifying wetting states}\label{Appendix_Phases}
  
We now discuss numerical methods for analysis of wetting transitions, in the presence of a barrier.
In simulations run for a sufficiently long time, we observe four distinct phases, described in detail below. However, obtaining the long-time limiting behaviour is numerically expensive so we instead developed a numerical method for classifying what will be the long-time behaviour, from finite simulations up to time $t_{\rm f}$. We verified the effectiveness of this procedure by performing long simulations at a few points on the surface phase diagram: it is generally accurate with errors in classification only appearing very close to the transition lines (usually at a distance less than the resolution of the surface phase diagram).

We choose as model parameters $(D_E, D_O, \bar{\rho}, L) = (1, 1, \bar{\rho}_\text{crit}, 12)$ and computational parameters $N_x = 500, N_\theta = 20, \tau_\text{max}  = 10^{-4}$. Repeating the analysis at single points with larger $N_x,N_\theta$ or smaller $\tau_\text{max}$ did not lead to any significant difference in temporal behaviour or spatial structure.
Our choices ensure that $L \gg l_D, w_b \gg \Delta x$, \textit{i.e.} the system is much larger than the typical scale of phase interfaces or the barrier, both of which are well resolved on the grid.

As an outline of the classification algorithm: States are classified first according to whether they are phase separated, then by whether the barrier is wet, and finally by whether the barrier is fully- or partially-wet.  Finally the wetting transition line is calculated. This we now describe.

\paragraph{Phase separated vs homogeneous states.}
The following variance of density away from its mean value is measured:
\begin{equation}
    \sigma^2_\rho =  \int_0^{L/2 - w_b} (\rho(x) - \bar{\rho})^2 \ \dd x + \int_{L/2 + w_b}^L (\rho(x) - \bar{\rho})^2 \ \dd x. 
\end{equation}
where we exclude the region $x \in [L/2 - w_b, L/2 + w_b]$ in order to remove the effect of volume exclusion on the barrier. States are classified as homogeneous if $\sigma^2_\rho < 0.01$. The blue phase-separation line shown in Figure \ref{fig:phase_behaviour}(a) is a smooth line fit to the data and of form $\text{Pe}_\text{PS} = \frac{a}{b+c \epsilon} + d$. 

\paragraph{Unpinned/dry vs wet states.}
As a barrier does not exist for $\epsilon = 0$, all states with $\epsilon=0$ are trivially dry/unpinned. To find whether the states at finite $\epsilon$ are pinned to the barrier, their density profiles are examined. If the barrier is within $l_D$ of a liquid-vapour interface or inside the liquid bulk, we consider the state to be wet. With this definition, all solutions at finite $\epsilon$ are classified as wet in steady-state. Experiments at smaller $\epsilon$ than the minimum value reported on Figure \ref{fig:phase_behaviour}(a) remain wet, indicating that any finite barrier height leads to wetting.

\paragraph{Partially-wet vs fully-wet states.}  Finally we settle the distinction between partially-and fully-wet:
states are classified as fully-wet if $\mathcal{A}<10^{-3}$ (i.e. the wetting films are approximately the same width); otherwise, they are classified as partially-wet.

\paragraph{Full-to-partial wetting transition.}
In order to generate the wetting line shown in Fig.~\ref{fig:phase_behaviour}(a), we make use of the linear instability of the fully-wet state. We measure the value of $\dot{\mathcal{A}}/\mathcal{A}$ at the time such that $\mathcal{A} = 10^{-3}$. A linear equation is then fit to these data-points of form $\dot{\mathcal{A}}/\mathcal{A} = a(\epsilon - \epsilon^*(\text{Pe}))$, using the 3 smallest values of $\dot{\mathcal{A}}/\mathcal{A}$ at a given Pe (3 values are taken in order to ensure the fit is within the linear instability regime). This gives an estimate for $\epsilon^*(\text{Pe})$, shown with the green wetting line in Figure \ref{fig:phase_behaviour}(a). 
The advantage of this method, compared to directly measuring the asymmetries and allocating states accordingly, is that we do not need to wait an exponentially long time for states near the transition to reach the partially-wet state. This therefore allows an accurate calculation of the wetting line with a reasonable computational cost. Comparing the wetting line generated with this method (i.e. measuring instabilities) with the above classification scheme for the steady-states (i.e. by measuring $\mathcal{A}$ directly), we see that only a small discrepancy is visible, as expected.

\bibliography{apssamp}

\end{document}